\documentclass[aps,prc,preprint,amsmath,showpacs,superscriptaddress,nofootinbib]{revtex4}
\usepackage{graphicx,epsfig}
\newcommand{\beqy}{\begin{eqnarray}}
\newcommand{\eeqy}{\end{eqnarray}}
\newcommand{\bmlet}{\begin{subequations}}
\newcommand{\emlet}{\end{subequations}}

\begin{document}

\textwidth 16.2 cm
\oddsidemargin -.54 cm
\evensidemargin -.54 cm

\def\gsimeq{\,\,\raise0.14em\hbox{$>$}\kern-0.76em\lower0.28em\hbox  
{$\sim$}\,\,}  
\def\lsimeq{\,\,\raise0.14em\hbox{$<$}\kern-0.76em\lower0.28em\hbox  
{$\sim$}\,\,}

\title{Breathing-mode measurements in Sn isotopes and isospin dependence of
nuclear incompressibility}
\author{J.~M.~Pearson}
\affiliation{D\'ept. de Physique, Universit\'e de Montr\'eal, Montr\'eal
(Qu\'ebec), H3C 3J7 Canada}
\author{N.~Chamel}
\affiliation{Institut d'Astronomie et d'Astrophysique, CP-226, Universit\'e
Libre de Bruxelles, 1050 Brussels, Belgium}
\author{S.~Goriely}
\affiliation{Institut d'Astronomie et d'Astrophysique, CP-226,
Universit\'e Libre de Bruxelles, 1050 Brussels, Belgium}
\date{\today}

\begin{abstract}
T. Li {\it et al.}[Phys. Rev. C {\bf 81}, 034309 (2010)] have analyzed their measured breathing-mode energies 
of some tin isotopes in terms of a first-order leptodermous expansion, and find
for the symmetry-incompressibility coefficient $K_{\tau}$ the value of 
-550 $\pm$ 100 MeV. Removing an approximation that they made, we find that 
the first-order estimate of $K_{\tau}$ shifts to -661 $\pm$ 144 MeV. However,
taking into account higher-order terms in the leptodermous expansion shows
that the data are compatible with the significantly lower magnitudes indicated 
by both another experiment and some theoretical estimates.

\end{abstract}


\maketitle

Li  {\it et al.}~\cite{lietal0} have measured the energies $E_{GMR}$ 
of the giant isoscalar monopole resonance in the seven even-even isotopes
$^{112-124}$Sn, expressing their results in terms of the finite-nucleus 
incompressibility 
\beqy\label{1}
K(Z, A) = \frac{M}{\hbar^2}R^2 E_{GMR}^2  \quad ,
\eeqy
where $R$ is the rms matter radius (see also Ref.~\cite{lietal}). Now adopting 
the leptodermous picture of 
the nucleus~\cite{ms69}, $K(Z, A)$ can be expanded about $K_v$, the 
incompressibility of symmetric nuclear matter, in powers of the small 
quantities $A^{-1/3}$ and $I^2$, where $I = (N - Z)/A$; if we retain only the 
lowest-order terms beyond $K_v$ we have~\cite{bla80}
\beqy\label{2}
K(Z, A) = K^{(1)}(Z, A) \equiv K_v + K_{sf}A^{-1/3} + K_{\tau}I^2
+ K_{coul}\frac{Z^2}{A^{4/3}} \quad .
\eeqy

However, as Refs.~\cite{lietal0,lietal} recall, it was shown long 
ago~\cite{pea91} that fitting Eq.~(\ref{2}) to the measured values of $K(Z, A)$
cannot determine a unique value of $K_v$, essentially because even though there
are only three other coefficients, the data are insufficiently accurate and too
few in number. Nevertheless, Refs.~\cite{lietal0,lietal} argue that it is still
possible to extract a value for $K_{\tau}$ from their data on the Sn isotopes. 
They do this by making three assumptions: i) the variation of $A^{-1/3}$ over 
the measured chain of Sn isotopes is so small that the first two terms on the 
right-hand side of Eq.~(\ref{2}) can be lumped together as a constant; ii) all
higher-order terms in the expansion (\ref{2}) can be neglected; iii) $K_{coul}$
lies within the range -5.2 $\pm$ 0.7 MeV. They then find that $K_{\tau}$
falls in the range -550 $\pm$ 100 MeV.

This result is to be compared with the value of -370 $\pm$ 120 MeV extracted
from measurements of isospin diffusion in heavy-ion collisions~\cite{chen09}. 
The two results are not inconsistent, but there is 
some theoretical support for a lower magnitude of $K_{\tau}$. For example, 
referring to Table 1 of Ref.~\cite{chen09} (where our $K_{\tau}$ is denoted 
by $K_{sat,2}$), we see that most Skyrme forces and various forms of 
Gogny forces predict values of $K_{\tau}$ lying in the range -325 $\pm$ 55 
MeV; the only exceptions in this table are forces for which $K_v$ is 
excessively large. A more recently developed Skyrme force, BSk18~\cite{cgp09},
which gives a precision fit to essentially all the mass data, and satisfies 
several realistic constraints, especially relating to neutron matter, has for
$K_{\tau}$ the value of -344 MeV. Actually, it is on the basis of this
disagreement between the value of $K_{\tau}$ that they extract from experiment
and the Skyrme values that the authors of Refs.~\cite{lietal0,lietal} conclude 
that their data
``rule out a vast majority of Skyrme forces". However, further theoretical 
support for a lower magnitude of $K_{\tau}$ comes in the form of three 
different Brueckner-Hartree-Fock calculations~\cite{vid09} which yield values 
of $K_{\tau}$ lying in the range -344 to -335 MeV. Moreover, and most 
significantly, it has been argued~\cite{piek09a,piek09b} that a value of -550 
MeV is too strongly negative
to be compatible with the behavior of low-density neutron matter, which is 
determined unambiguously by low-energy neutron-neutron scattering.

Accordingly, in this note we revisit the analysis of 
Refs.~\cite{lietal0,lietal} in an
attempt to see whether their quoted error bars could be widened to accommodate 
the lower magnitude of $K_{\tau}$ for which there appears to be significant 
evidence. We find their assumption (iii) (see above) to be reasonable but 
contest assumptions (i) and (ii), as follows.

{\it Non-constancy of $A^{-1/3}$.} If we take account of the variation of 
$A^{-1/3}$ over the measured chain of Sn isotopes then fitting Eq.~(\ref{2}) to
the data will require assumptions about $K_v$ and $K_{sf}$. For the former we
take the well established range 240 $\pm$ 10 MeV~\cite{col04}, while $K_{sf}$ is
used along with $K_{\tau}$ to fit the data (the fits of 
Refs.~\cite{lietal0,lietal} had 
only one free parameter, $K_{\tau}$). Taking into account the range of 
uncertainty on $K_{coul}$ assumed by Ref.~\cite{lietal}, we find that fitting
Eq.~(\ref{2}) to the data yields $K_{\tau}$ = -661 $\pm$ 144 MeV. These lower
and upper limits of $K_{\tau}$ correspond to Sets 1 and 2 of Table I and curves
1 and 2, respectively, in Fig. 1. 

Comparison of this new value for $K_{\tau}$ with the value of 
-550 $\pm$ 100 MeV given by Refs.~\cite{lietal0,lietal} shows that the 
variation 
of $A^{-1/3}$ over the isotope chain is significant, even if it amounts to only
3\%. More seriously, our new analysis of the data has aggravated the conflict 
with both theory and the only other recent measurement of 
$K_{\tau}$~\cite{chen09}.
Nevertheless, we now show that it is possible to reconcile the data with a
significantly smaller magnitude of $K_{\tau}$: 
bearing in mind the values indicated by the heavy-ion experiment~\cite{chen09}
and the various theoretical calculations mentioned above, we shall, to be 
specific, consider just the value of -350 MeV.
  
{\it Higher-order terms.} Going to the next order in powers of $A^{-1/3}$ and 
$I^2$, Eq.~(\ref{2}) is replaced by
\beqy\label{3}
K(Z, A) = K^{(1)}(Z, A) + K_{ss}I^2A^{-1/3} + K_{cv}A^{-2/3} + K_{\tau4}I^4
\quad .
\eeqy
Of the three new terms, the last one relates to infinite nuclear matter, whence
the coefficient $K_{\tau4}$ can be reliably calculated for Skyrme forces. 
However, Table~I of Ref.~\cite{chen09} (where our $K_{\tau4}$ is denoted by
$K_{sat,4}$) makes it clear that this term will make only a negligible 
contribution to $K(Z, A)$, so we study the effect only of the $K_{ss}$ 
(surface-symmetry) and $K_{cv}$ (curvature) terms (a possible role for the 
surface-symmetry term was suggested by Col\`o~\cite{col09}). Setting $K_v$ = 
240 MeV, $K_{coul}$ = -5.2 MeV and $K_{\tau}$ = -350 MeV, we leave $K_{sf}$,
$K_{ss}$ and $K_{cv}$ as fitting parameters. Since only two parameters are
required for a good fit to the data, an infinite number of fits are possible, 
one for each value of $K_{cv}$. We took just two values for $K_{cv}$, simply
to assess the role of the curvature term, and defined thereby Sets 3 and 4 of
Table I; the corresponding curves in Fig.~I are indistinguishable from the 
curve 2, corresponding to Set 2. 

{\it Conclusions.} We see that by invoking the surface-symmetry term it is
possible to reconcile the breathing-mode data of Refs.~\cite{lietal0,lietal} 
with a value of -350 MeV for $K_{\tau}$; this value is consistent with
the heavy-ion experiment of Ref.~\cite{chen09} and the various theoretical
estimates that we have mentioned above. The required value of $K_{ss}$ can be 
reduced if we make use of the curvature term as well. Of course, one might ask 
whether the required values of $K_{sf}$, $K_{ss}$ and $K_{cv}$ are physically 
plausible. In principle, we could answer this question by making the appropriate
calculation of semi-infinite nuclear matter with various forces, but such
calculations tend to be somewhat unreliable because of stability problems.
In any case, our required values of $K_{sf}$, $K_{ss}$ and $K_{cv}$ must be
regarded as ``effective" values, since they absorb the effect of all the
terms of still higher order, such as a term in $I^2A^{-2/3}$, that we have 
neglected in fitting the data. In fact, it has been suggested~\cite{trei86} 
that the leptodermous expansion may not converge at all, no matter how many 
terms are taken.The only reliable approach to the problem would be to 
calculate the breathing-mode energies of each nucleus directly through 
self-consistent quasiparticle random-phase approximation (QRPA) (or constrained 
Hartree-Fock-Bogoliubov (HFB)) calculations with a succession of 
different forces until agreement with the data is 
reached~\cite{bla95}. This is quite beyond the scope of the present note. 

This state of affairs is somewhat unsatisfactory. However, we remark that
the situation is little better with the analysis of 
Refs.~\cite{lietal0,lietal}: the
tacit assumption of zero values for $K_{ss}$ and $K_{cv}$ is just as much in 
need of justification as are our required non-zero values. Thus it was 
altogether premature of Refs.~\cite{lietal0,lietal} to conclude that their data
``rule out a vast majority of Skyrme forces". 

We thank U. Garg and G. Col\`o for correspondence. 
The financial support of the NSERC (Canada), 
the Communaut\'e fran\c{c}aise de Belgique (Actions de Recherche 
Concert\'ees) and the FNRS (Belgium) is
acknowledged. J. M. P. is grateful to the Bureau des
relations internationales of the Universit\'e Libre de Bruxelles for
financial support during the month of November 2009.

\newpage

\begin{table}
\centering
\caption{Parameter sets (in MeV) for calculating $K(Z, A)$.} 
\label{tab1}
\vspace{.5cm}
\begin{tabular}{|c|cccccc|}
\hline
& $K_v$ & $K_{coul}$ & $K_{\tau}$ & $K_{sf}$ & $K_{ss}$ & $K_{cv}$\\
\hline
Set 1& 250.0 & -5.9 & -805.0 & -386.0 & 0 & 0 \\
Set 2& 230.0 & -4.5 & -517.0 & -352.0  & 0 & 0 \\
Set 3& 240.0 & -5.2 & -350.0 & -382.5  & -980.0 & 0 \\
Set 4& 240.0 & -5.2 & -350.0 & -488.0  & -835.0 & 500.0 \\
\hline
\end{tabular}
\end{table}

\vspace{5.0cm}

\begin{figure}
\centerline{\epsfig{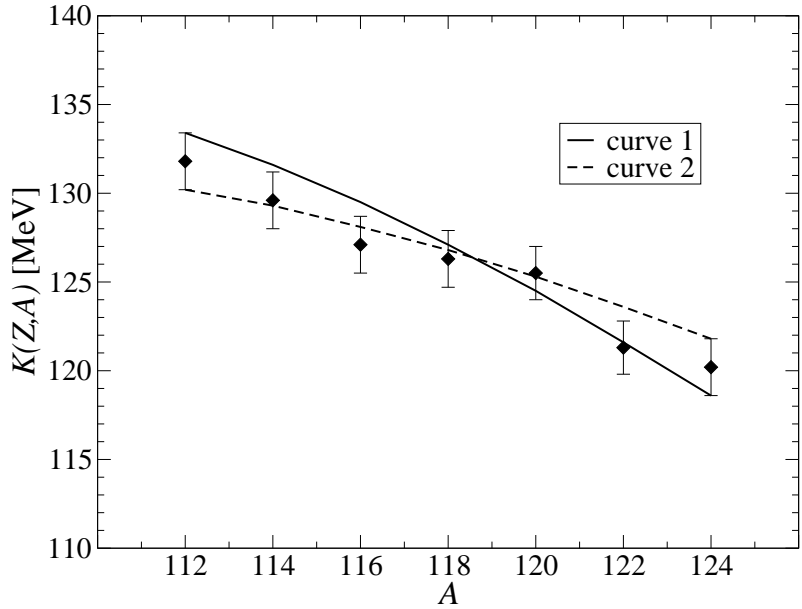}}
\caption{Fits of parameter sets of Table~\ref{tab1} to experimental values of
$K(Z, A)$. Curve 1 corresponds to Set 1, curve 2 to Sets 2, 3 and 4.} 
\label{fig1}
\end{figure}
\end{document}